\documentclass[journal]{IEEEtran}
\usepackage{graphicx}
\usepackage{booktabs}
\usepackage{array}
\usepackage{url}
\usepackage[T1]{fontenc}
\usepackage[utf8]{inputenc}
\usepackage{textcomp}
\usepackage{amsmath}
\usepackage{amssymb}
\graphicspath{{figures/}}

\begin{document}

\title{Where to Decide: Control-Plane Geometry in\\Coherence-Limited Quantum Networks}

\author{Indrakshi~Dey,~\IEEEmembership{Senior Member,~IEEE} and Nicola~Marchetti,~\IEEEmembership{Senior Member,~IEEE}%
\thanks{I.~Dey is with Department of Computing and Mathematics, South East Technological University, Waterford, Ireland; N.~Marchetti is with Department of Electrical and Electronic Engineering, Trinity College Dublin, Dublin, Ireland (e-mail: indrakshi.dey@setu.ie; nicola.marchetti@tcd.ie). This work is supported by HORIZON MSCA-DN Project ``QUESTING" under Grant Number 101227218. \emph{This paper is under review at IEEE Network, Series on Quantum Communications and Networking.}}}

\markboth{IEEE Network, Series on Quantum Communications and Networking}%
{Where to Decide: Control-Plane Geometry in Coherence-Limited Quantum Networks}

\maketitle

\begin{abstract}
Quantum networks are usually framed by two hardware limits: how fast entanglement
can be heralded, and how long a memory can hold it. We establish a third,
geometric limit: entanglement decoheres while control information travels, so the
distance to whoever allocates resources enters the fidelity budget directly. We
develop a functional-graph abstraction weighting each link by its predicted
multiplicative contribution to end-to-end fidelity, separating global but delayed
state from local and immediate state, and compare centralized against local
allocation in a replicated discrete-event model. Centralized delivery latency
grows with network diameter while local latency is nearly scale invariant,
differing up to twenty-fold for short-range traffic; the coherence threshold at
which a global view outweighs a fresh one does not shift across the tested sizes;
and as heralding accelerates, the dominant cost moves into the control plane.
Slot synchronization instead obeys the product of slot rate and one-way link
delay, so it is bounded by repeater spacing rather than diameter. Controller
placement, decision locality, repeater spacing and slot rate are physical design
parameters, not implementation details.
\end{abstract}


\section{Introduction}

\IEEEPARstart{A}{quantum} network distributes entanglement. Two distant nodes
share a Bell pair, which they consume for key distribution, distributed quantum
computation, or quantum-enhanced sensing \cite{kimble2008, wehner2018}. Between
them sit repeaters that generate entanglement on short elementary links and
splice those links by entanglement swapping \cite{briegel1998}; the resulting
architecture and protocol stack are now the subject of serious engineering
effort \cite{vanmeter2013, pirker2019}. Almost every engineering discussion of
these networks concerns hardware: source brightness, fiber loss, memory
coherence. These are the right questions, but not the only ones.

A second limit is geometric. Entanglement is perishable: a pair decays from the
moment it exists, and must be held until every other segment of the path is ready
before the swaps can proceed. Anything delaying assembly of the chain is
therefore not a latency cost but a fidelity cost, paid in the same currency as
fiber attenuation and memory noise. One such delay is the round trip to whichever
entity allocates resources. In a classical network a controller taking a
millisecond makes the network slower, not incorrect; in a quantum network that
millisecond is subtracted from a coherence budget that may itself be a
millisecond long, and the controller's position becomes a physical parameter.

Entanglement distribution has been approached from several directions: routing
metrics for probabilistic heralding \cite{pant2019}, link-layer protocols
exposing entanglement as a service \cite{dahlberg2019}, flow-based rate
characterizations \cite{chakraborty2020}, distributed routing
\cite{chakraborty2019}, and concurrent-request scheduling \cite{shi2020}. These
share an implicit assumption that the information on which an allocation decision
rests is available when the decision is made. In a quantum network it is not,
because the resource being allocated expires while information about it is in
transit.

A second, quieter problem sits alongside. Swapping requires the two adjacent links
to attempt heralding in the same slot, and without a common clock, agreeing on
that slot is itself a distributed task subject to the same finite signaling
speed. Timing and routing are usually designed as separate layers, and their
interaction has received little attention.

\begin{figure*}[!t]
\centering
\includegraphics[width=0.8\linewidth]{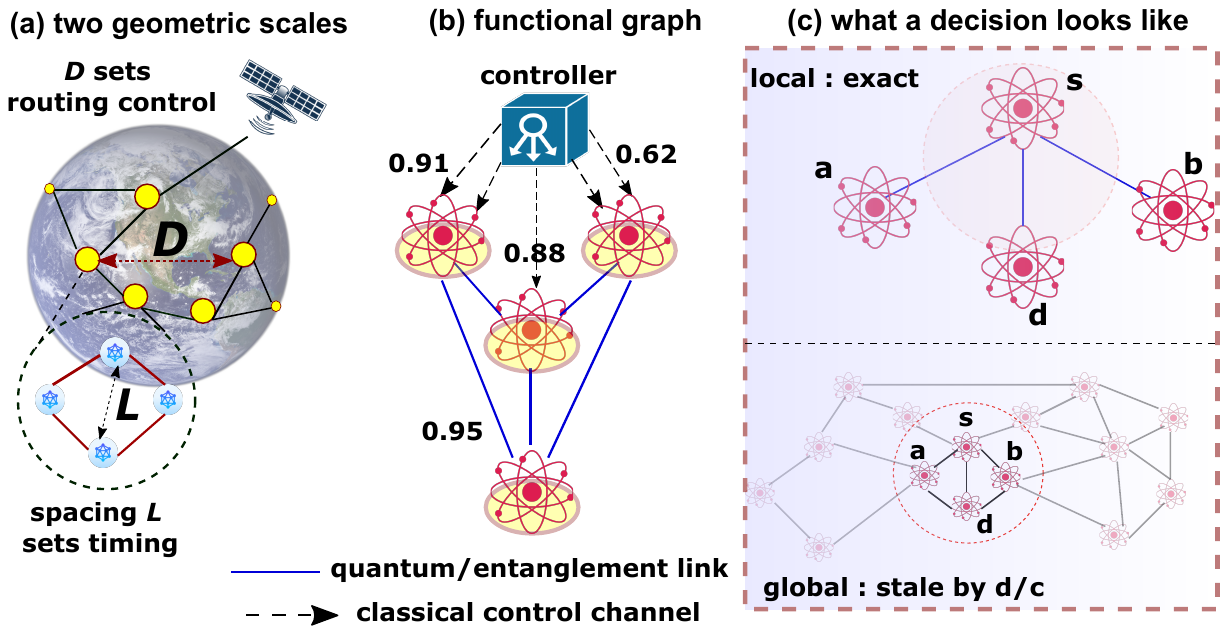}
\vspace{-133mm}
\caption{(a) The two geometric scales. Timing coordination between adjacent
repeaters is set by the spacing $L$; routing control is set by the diameter $D$,
because every request pays a round trip to the controller however short its own
path. (b) The functional graph: each edge weight is the predicted contribution
of that link to end-to-end fidelity, combining how good a pair is when created,
how long it must be waited for, and how fast it decays. Thicker edges carry
higher weight, and the best path is the one whose weights multiply to the
largest value. (c) The information behind one decision: a node sees its own
incident links exactly, a controller sees all links but each as it was one
propagation delay ago. In (b) and (c) the circled letters name network nodes,
$s$ the source and $d$ the destination; the $d$ appearing in the staleness
$d/c$ of (c) is a different quantity, the node-to-controller distance of
Section~\ref{sec:fidloss}. Provenance: schematic, no measured or computed data.}
\label{fig:concept}
\end{figure*}

In this paper we treat, for the first time, the control plane as a physical
subsystem whose geometry enters the fidelity budget. We develop a
functional-graph abstraction weighting each link by its predicted multiplicative
contribution to end-to-end fidelity, separating quasi-static topology from
rapidly changing dynamic state. We then establish that coordination obeys two
distinct geometric limits: routing and allocation are constrained by network
diameter, distributed slot synchronization by neighboring-repeater spacing.
Centralized and local coordination are therefore not alternative implementations
of one function but are bounded by different physical scales.

\section{The Functional Graph}

\subsection{Why Decision Time Becomes Fidelity Loss}
\label{sec:fidloss}

Consider a path of several elementary links. A heralded pair has an initial
quality representable by a depolarization parameter that decays exponentially in
memory; under ideal swapping these parameters multiply, so end-to-end quality is
the product of the surviving link and swap qualities. This makes waiting
physically important: a pair created early cannot be consumed while another
segment is missing, and every additional delay enters the same exponential decay
as the hardware imperfections: finite source visibility and residual
multi-photon emission, detector dark counts, non-unit Bell-state measurement
fidelity at each swap, and memory dephasing and read-out error. Holding a pair
for 100\,\textmu s in a memory with 1\,ms coherence costs that segment roughly
ten per cent. The delay changes the resource the quantum process produces; it is
not an overhead attached afterwards.

Two distances create this waiting (Fig.~\ref{fig:concept}(a)). Under centralized
allocation a request travels to the controller and the instruction travels back,
a round trip of $2d/c$ in which $d$ is the fiber distance from the requesting
node to the controller and $c$ the signal speed in fiber. That round trip scales
with network diameter even when the communicating users are adjacent, and the
one-way term $d/c$ sets the age of the dynamic state the controller holds when
it decides. Reference-free slot alignment, by contrast, signals only between neighboring repeaters, so its penalty is set by elementary-link length. Treating both as generic control latency hides the distinction: one grows with the extent of the network, the other is fixed by its local spacing.

\subsection{Weights in the units of delivered fidelity}

Classical routing represents a link by capacity, cost or delay and combines those additively along a path. An entanglement link carries no durable flow but generates a perishable resource at random times, so its usefulness depends jointly on the quality of a newly heralded pair, the rate at which pairs arrive, and the coherence time over which they stay usable. The interaction extends beyond the individual edge. A slow link does not merely
delay its own contribution; it forces already prepared segments elsewhere to wait
and decohere, so the cost of one edge is paid across the chain, which an additive
capacity metric cannot represent. We model the network as a graph whose vertices
hold memories and whose edges are elementary links, each carrying a weight formed
from the quality of a freshly heralded pair and an exponential penalty for the
expected wait until a usable pair exists, scaled by coherence time. That penalty
is zero when a suitable pair is buffered and follows the mean heralding interval
when the buffer is empty.

The resulting weight is neither a raw probability nor a conventional delay or
capacity: it predicts that edge's multiplicative contribution to the fidelity of
any chain using it. Maximizing the product of edge weights therefore optimizes
what the network realizes rather than a surrogate, and taking logarithms converts
that product into an additive path cost, so conventional shortest-path algorithms
operate on the functional graph without altering the objective. The weights of
Fig.~\ref{fig:concept}(b) show why the product rather than the sum is the right
combination: the route $s\to a\to d$ scores $0.91\times0.88=0.80$ against
$0.62\times0.95=0.59$ for $s\to b\to d$, so the strongest single edge in that
graph lies on the losing path. One weak link cannot be redeemed by strong
remaining links. Fidelity-weighted edges and logarithmic path costs have
precedents: an end-to-end entanglement metric minimized by a Dijkstra-type
algorithm \cite{caleffi2017}, and rate-oriented metrics for probabilistic
heralding \cite{pant2019}. The distinctive step here is to connect those weights
to the age and location of the information from which they are evaluated.

\subsection{Topology and dynamic state are different knowledge}

A functional weight can guide a decision only if the decision maker can evaluate
it, which exposes a distinction classical networking has little reason to make.
Topology (which nodes and links exist, how they connect, how long the fibers
are) changes slowly, so the network can distribute it globally and keep it
consistent. Dynamic entanglement state changes far faster: which links hold pairs, their ages
and qualities, the memory free at each node. A remote controller receives this
only as a report delayed by propagation and processing, by which time some pairs
have decayed, been consumed or been replaced, and no protocol can remove the
propagation component of that age. The same functional graph therefore has two
legitimate readings (Fig.~\ref{fig:concept}(c)): a node sees only its incident
links, but accurately and now, while a controller sees the entire graph with each
part at a different time in the past. The choice is not between limited and
complete information, but between narrow, fresh and broad, stale information.

\subsection{Comparing decision locality without changing the objective}

To isolate this tradeoff, both schemes share the same graph, hardware and fidelity
objective. The centralized scheduler precomputes candidate fewest-hop paths from
the quasi-static topology and scores them by the product of its delayed
observations of the dynamic weights. It is intentionally strong (global scope,
evaluating the exact predicted objective), and its only disadvantage is that
every request triggers a remote decision on aged state. The local scheduler
forwards hop by hop, combining exact knowledge of a node's incident weights with
a quasi-static distance table and selecting the neighbor offering the best
fidelity-weighted progress. This rule is deliberately naive, which is useful:
where local operation stays competitive, the advantage comes from freshness and
locality, not from a better algorithm.

\section{Simulation Setup}
\label{sec:setup}

The discrete-event model represents a metropolitan repeater network over a square
service area with nearest-neighbor links. Unless varied explicitly, mean repeater
spacing is approximately 17\,km. Elementary links herald pairs as Poisson
processes whose rates follow from fiber attenuation at telecom wavelength, source
brightness and detector efficiency. Each node stores a few pairs per incident
link. A stored pair is discarded once its fidelity decays below the delivery
threshold $F_{\mathrm{th}}$, and a chain counts toward goodput only if the
delivered pair still exceeds that threshold, so goodput measures useful
deliveries rather than attempts. Requests arrive as a Poisson process with destinations drawn from a distance-decay
distribution whose characteristic distance is the traffic-locality scale $\ell$:
selection probability falls off with separation on the scale $\ell$, so small
$\ell$ means demand concentrated among near neighbors and large $\ell$ means
demand spread across the network. Once $\ell$ approaches the diameter the decay
stops discriminating and traffic is effectively uniform. The controller sits at
the node nearest the geometric centroid, its most favorable location.

The assumed brightness and detector efficiency give a few thousand pairs per
second over a 17\,km link, well above current demonstrations and presuming future
multiplexed sources \cite{humphreys2018}. The choice is deliberate: when
heralding is overwhelmingly slow, physical generation hides the control cost this
paper examines. The control shares in Section~\ref{sec:budget} therefore describe
a future fast-heralding regime, not today's operating point, and scheduler
comparisons remain the principal evidence since both face the same assumptions.
Each reported allocation point (one combination of diameter, locality scale and
coherence time for one scheduler) averages six replicates with separate
topologies, arrivals and heralding seeds, shown with ninety-five per cent
confidence intervals. Replication matters because individual runs differ by about
five per cent.

\subsection{Remote allocation inherits network diameter}

\begin{figure*}[!t]
\centering
\includegraphics[width=\linewidth]{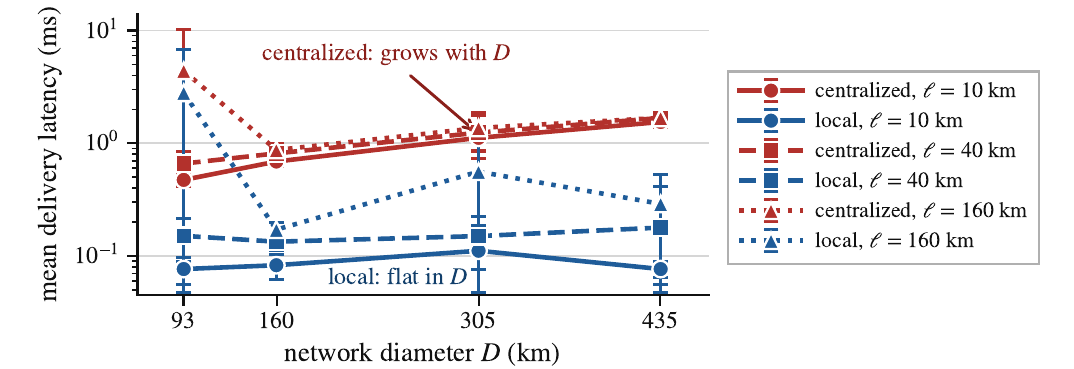}
\caption{Mean delivery latency against network diameter at constant node density, for
three traffic-locality scales $\ell$. Centralized latency grows with diameter;
local latency is nearly scale invariant. Six replicates per point, ninety-five
per cent confidence intervals, 1\,ms coherence.}
\label{fig:latency}
\end{figure*}

Figure~\ref{fig:latency} varies network diameter at fixed node density, and therefore fixed repeater spacing, for short-, medium- and long-range traffic. For short- and medium-range demand centralized latency rises with diameter: under short-range traffic from 0.47\,ms at 93\,km to 1.55\,ms at 435\,km. Local latency stays between 0.08\,ms and 0.18\,ms across the same fivefold increase in scale. The smallest network carrying the most dispersed traffic falls outside this
pattern, for a reason worth stating precisely. A locality scale of 160\,km inside
a 93\,km network means $\ell$ exceeds the diameter, so the distance decay never
engages and every node is nearly equally likely as a destination. Almost every
request then spans the network, few paths are feasible, and a handful of very
long waits dominate the mean, producing the wide confidence interval at the left
of Fig.~\ref{fig:latency}. That point reflects a request pattern the network is
too small to contain, not a reversal of the trend. Outside this regime the
centralized-to-local ratio reaches 20.1 for short-range traffic in the largest
network and lies between 1.6 and 5.8 for dispersed traffic.

The mechanism matters more than the ratio. A central controller charges even a
local request for a round trip to the decision locus, so the penalty grows as
infrastructure expands, whereas local decisions stop the fidelity cost of
authorization compounding with diameter. Two lessons follow. First, the cost of
centralization is set by the ratio of locality scale to diameter rather than by
network size, since a short path pays the same round trip as a long one, so
decision locality should be chosen from the demand profile rather than the node
count. Second, the penalty is largest where demand is dense and short-range: a
network growing outward while its traffic stays local is where centralized
control degrades fastest.

\subsection{Memory determines when global optimization is worth stale state}

\begin{figure*}[!t]
\centering
\includegraphics[width=\linewidth]{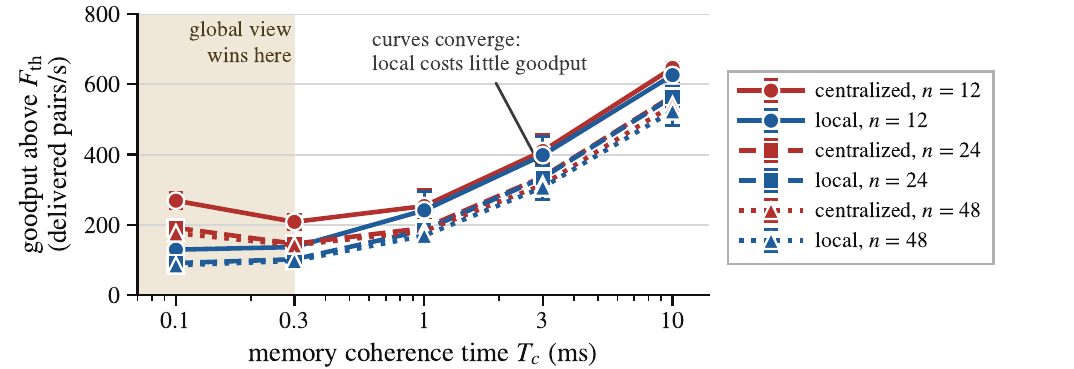}
\caption{Goodput against memory coherence time in networks of 12, 24 and 48 nodes at fixed
node density, so service area and diameter grow with node count. Goodput counts
only pairs delivered above the threshold $F_{\mathrm{th}}$ of
Section~\ref{sec:setup}. The global view wins under short coherence (shaded
band), and the crossover stays in the same range at all sizes.}
\label{fig:coherence}
\end{figure*}

One might expect local decisions to matter most when memories are short lived,
since fresh state should then be at a premium. Figure~\ref{fig:coherence} shows
the opposite. Below roughly 300\,\textmu s the centralized scheduler delivers
higher goodput, at 100\,\textmu s roughly twice that of the local rule with
non-overlapping confidence intervals; above 1\,ms the two are statistically
indistinguishable. When coherence is very limited, freshness cannot repair a poor
path choice: the greedy local rule sometimes commits to a chain a global
optimizer would reject, and the memory leaves too little margin to absorb that
mistake. As coherence grows, the penalty of an occasional bad local decision
falls while the latency advantage of avoiding the round trip remains.

More importantly, the crossover does not move detectably across networks of 12,
24 and 48 nodes. Node density is fixed in all three, at the same mean spacing of
approximately 17\,km, so service area and diameter grow with node count, from 93
to 305\,km: the comparison varies geographic extent rather than packing more
nodes into a fixed footprint. The ratio of local to centralized goodput is 0.48
at 100\,\textmu s, 0.66--0.70 at 300\,\textmu s, and 0.89--0.95 at 1\,ms.
Three sizes cannot establish universal invariance, but the stability suggests the
transition is controlled by the ratio between memory coherence and the heralding
interval of an elementary link, both local hardware properties. If this persists
under broader models, a control paradigm could be selected from a bench
characterization of memory and link rather than recomputed for every deployment
scale.

\subsection{Faster hardware can move the bottleneck into the control plane}
\label{sec:budget}

\begin{figure*}[!t]
\centering
\includegraphics[width=\linewidth]{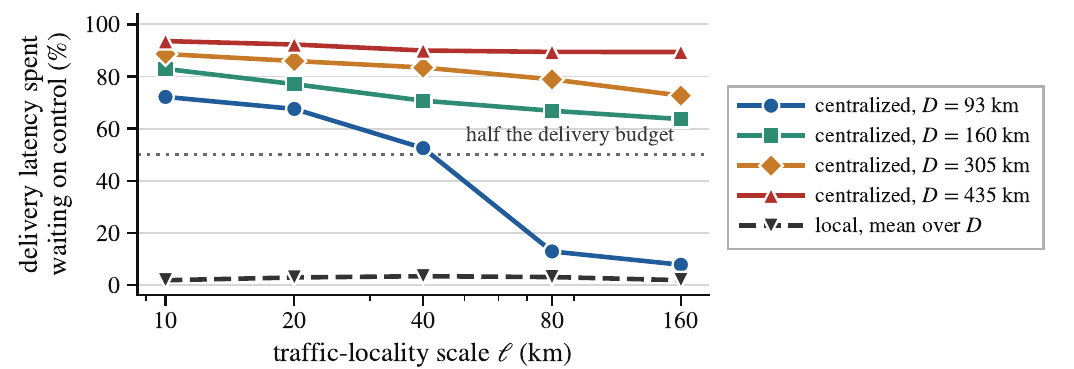}
\caption{Fraction of delivery latency spent waiting on control rather than physical
generation, against traffic-locality scale $\ell$, for four centralized diameters
and for local allocation averaged over all diameters. Conditional on the fast-heralding regime;
six replicates per point at 1\,ms coherence.}
\label{fig:budget}
\end{figure*}

Delivery latency separates into the time to authorize and activate a chain and
the time waiting for heralding and swap-outcome signaling. The simulator
timestamps three events per request: $t_0$, when the request is issued; $t_1$,
when the allocation decision reaches the chain's last node; and $t_2$, when that
node's final link is heralded and swapped. Those three instants bracket two
consecutive intervals, and the separation above is the split between them. The
first, $t_1-t_0$, is the control component, since no quantum resource can be
committed until the decision arrives. The second, $t_2-t_1$, is the physical
one, spent on heralding and swap-outcome signaling. Being consecutive they are
disjoint and together span the whole request, $t_2-t_0$, so the control share in
Fig.~\ref{fig:budget} is a ratio of measured intervals rather than a quantity
inferred from a model of the controller. In the 435\,km network
centralized control takes 89--94 per cent of delivery latency across the locality
range, against 73--89 per cent at 305\,km and 64--83 per cent at 160\,km; local
allocation spends 0.2--5.5 per cent, mean 2.5 per cent. The centralized share
falls to 8 per cent only in the smallest network under the most dispersed
traffic, where heralding dominates.

Two qualifications apply. First, these shares depend on the optimistic heralding
regime above; at present laboratory rates generation would occupy a larger share,
so the result identifies where the bottleneck can move as sources improve, not
that control dominates today's experiments. Second, the centralized scheduler is
reactive; proactive allocation avoids many round trips when demand is
predictable, but cannot eliminate the first decision for an unexpected
destination. Subject to that scope, in the 435\,km case doubling source
brightness targets roughly six per cent of the latency, whereas moving or
distributing the decision function addresses the remaining ninety-four. Hardware
progress does not remove the geometric cost of control; it makes that cost
visible.

\subsection{Reference-free timing follows repeater spacing}

Routing is not the only decision process affected by propagation: adjacent links
must attempt heralding in compatible slots for swapping to proceed. Deployments
with a stable out-of-band reference, such as GNSS-disciplined clocks or a White
Rabbit layer, coordinate externally, so the limit below applies to
reference-denied operation or to fallback after that reference is lost. Both are
ordinary: metropolitan fiber runs through tunnels and basements where a GNSS
antenna cannot be sited; GNSS is subject to jamming and spoofing; and a network
distributing time over fiber loses it whenever the grandmaster fails or a cut
isolates a segment. We represent each node as a slot oscillator coupled to its
graph neighbors, following the Kuramoto framework \cite{kuramoto1984}, which has
also been connected to distributed quantum systems \cite{komar2014}. Coupling
evolves over seconds whereas propagation takes tens of microseconds, so the delay
becomes a fixed phase lag \cite{sakaguchi1986}. Around cycles these lags cannot
generally sum to zero, so no phases align every edge once the propagation
fraction grows too large.

\begin{figure*}[!t]
\centering
\includegraphics[width=\linewidth]{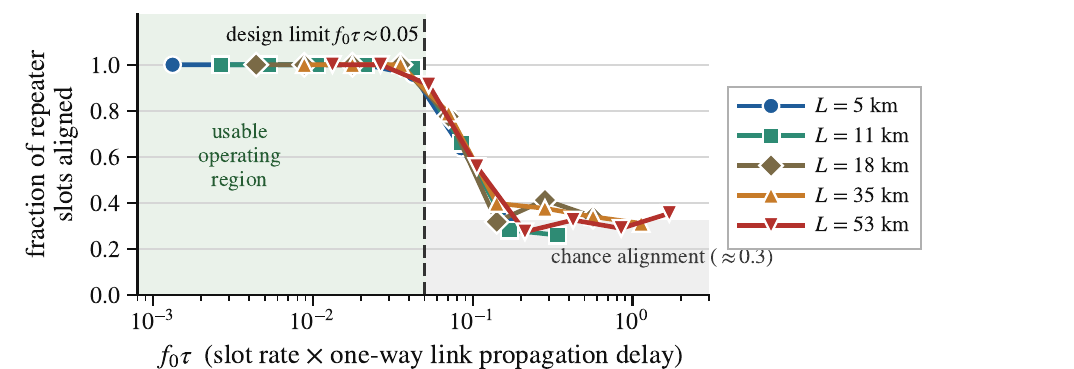}
\caption{Fraction of aligned repeater slots against the product of slot rate $f_0$ and
one-way link delay $\tau$, for five repeater spacings $L$. The collapse onto one
relationship shows that reference-free timing is governed by the local fraction
$f_0\tau$, not by network diameter. The shaded band marks chance alignment.
Three seeds per point, 10\,ppm clock stability.}
\label{fig:sync}
\end{figure*}

Figure~\ref{fig:sync} plots the fraction of aligned repeater slots against the
product of slot rate and link delay, for spacings from 5 to 53\,km. Despite the
tenfold range in distance, all five curves collapse onto one relationship:
alignment is almost perfect below about 0.035, holds between 0.90 and 0.93 near
0.05, and falls to chance beyond about 0.15. Chance is the alignment
independently and uniformly distributed phases would produce, approximately 0.3
for the slot-overlap tolerance used;\footnote{Once coupling has ceased to
organize them, the repeaters' slot phases are independent and uniform over the
slot period. Two neighbors then count as aligned whenever their circular phase
difference falls inside the overlap tolerance, which spans a fraction $w$ of the
period on either side of coincidence, an event of probability $2w$ for
$w\le\tfrac12$. The tolerance used throughout the sweep, $w\approx0.15$, puts
the floor at $\approx0.3$. It follows from $w$ alone and not from $L$ or $f_0$,
which is why a single horizontal band in Fig.~\ref{fig:sync} serves all five
spacings.}
reaching it means adjacent repeaters open
their windows together only by coincidence, so the coupling yields no usable
swapping schedule. Raising the coupling gain does not help, because the
obstruction is incompatible phase lags around the topology rather than
insufficient control authority.

The collapse gives a simple design rule. Write $\tau=L/v$ for the one-way delay
over a link of length $L$, with $v\approx2\times10^8$\,m/s the group velocity in
fiber. The abscissa $f_0\tau$ is then that delay expressed as a fraction of a
slot period, which is what makes it the natural variable: it is the phase lag
one edge imposes, measured in cycles, and the curves collapse because alignment
depends on the lag and not separately on the distance or the rate that produced
it. Alignment survives while $f_0\tau\lesssim0.05$, so the lag must stay under
about a twentieth of a cycle, equivalently the slot period must be at least
about twenty times $\tau$. Past that margin each edge carries an appreciable
fraction of a cycle, and since these lags cannot generally sum to zero around a
closed cycle of the topology, no assignment of phases can satisfy every edge at
once and alignment falls toward the chance floor.

Reading the rule as a rate bound, $f_0\lesssim0.05/\tau$. The mean spacing used
here gives $\tau\approx88$\,\textmu s and so $f_0\lesssim570$\,Hz: the
reference-free ceiling is just under 600\,Hz.
That ceiling constrains the rate at which the network can agree to open
windows, not the rate a node can attempt locally: a repeater may fire far faster,
but an attempt outside its neighbor's window has no partner to swap with and is
wasted. Unlike routing control, this ceiling does not grow with diameter: a
large dense network can be easy to synchronize yet poorly served by one distant
controller, and a small sparse network the reverse.

\section{Positioning and Design Implications}

Table~\ref{tab:sota} positions the two decision models against published
approaches on the assumptions that matter here: where a decision is made, what
age of dynamic state it assumes, and whether the delay in obtaining that state is
charged against coherence. It is not a performance benchmark, since no common
configuration exists, but a modeling boundary. Prior work has largely optimized
allocation while treating acquisition of the required state as free: reasonable
when the resource persists, consequential when it decays while its state report
travels.

\begin{table*}[!t]
\caption{Control-plane assumptions of representative entanglement-distribution approaches.
Entries describe each cited work and are not measured comparisons.}
\label{tab:sota}
\centering
\renewcommand{\arraystretch}{1.2}
\begin{tabular}{@{}p{3.2cm}p{2.6cm}p{2.9cm}p{2.6cm}p{4.2cm}@{}}
\toprule
\textbf{Approach} & \textbf{Decision locus} & \textbf{Dynamic state assumed} &
\textbf{Control delay charged} & \textbf{Primary reported quantity} \\
\midrule
Routing, flow and concurrent-request optimization \cite{pant2019},
\cite{chakraborty2020}, \cite{shi2020} & Global & Instantaneous & No &
Rate, bounds, throughput \\
Link layer protocol \cite{dahlberg2019} & Per link & Local, current & Partly,
within the link & Service abstraction, demonstration \\
Distributed routing \cite{chakraborty2019} & Distributed & Local & No &
Path quality, reachability \\
This work, centralized & Global & Delayed by $d/c$ & Yes &
Latency, goodput, control share \\
This work, local & Per node & Local, exact & Yes &
Latency, goodput, control share \\
\bottomrule
\end{tabular}
\end{table*}

We do not claim a superior routing algorithm; the local rule is intentionally
naive and is beaten by the global optimizer when coherence is very short. The
contribution is the accounting that makes the control plane draw on the same
finite budget as the quantum resource, and the resulting separation of global
decision distance from local timing distance.

Controller placement belongs in physical network planning. The relevant distance
is not the controller's distance to the most remote node but the detour between
the controller and the paths carrying typical demand, so a controller at the
geometric center may still sit far from predominantly local traffic.
Regionalization shortens $d$ in the $d/c$ penalty: a controller serving a 40\,km
region charges a round trip across that region rather than the network.
Pre-allocation leaves $d$ unchanged but amortizes it, since one round trip can
authorize a standing allocation serving many later requests. Neither helps the
first request to an unanticipated destination.

The control paradigm should also match the memory technology. Coherence above
roughly 1\,ms lets local decisions retain goodput while giving a large,
scale-invariant latency benefit; below a few hundred microseconds, global path
quality outweighs fresh local state. The crossover is stable in one specific
sense: it does not shift measurably across the three sizes simulated, 12 to 48
nodes and 93 to 305\,km. That is what would let bench measurements of coherence
and heralding rate guide the choice before the deployment size is known. It is
not a claim of stability with respect to topology family, traffic model or
protocol.

Finally, the slot rate must be selected from the repeater spacing when no external
timing reference can be assumed, and the gap this opens is large. Consider a
source able to attempt generation five thousand times per second. Over links with
$\tau \approx 88$\,\textmu s, the rule caps the coordinated rate at just
under 600\,Hz. That is the network-wide rate at which every repeater agrees to
open its attempt window, so two adjacent links attempt in the same slot. Roughly nine of
every ten attempts the source can make therefore cannot be placed in a slot the
neighbor is also using, and are wasted. Three remedies exist, each with a cost:
lower the coordinated rate and accept the throughput loss; shorten the spacing,
reducing $\tau$ but requiring more repeater sites; or supply an out-of-band
reference and accept the dependence. The datasheet attempt rate is a local
capability of the source alone, whereas the usable reference-free rate is the
smaller of two quantities: that attempt rate, and the coordination ceiling
$0.05/\tau$ fixed by the repeater spacing. In the example above the second binds,
5000\,Hz of local capability against a 570\,Hz ceiling. Because one is a source
property and the other a property of network geometry, neither can be inferred
from the other, and the two belong on separate specification lines.

These implications point to architectures between the two extremes: a hierarchical
controller combining regional freshness with broader path optimization, or
proactive allocation preserving a global view for predictable demand while
delegating unexpected requests locally. Purification and multiplexing would shift
the numerical boundaries but neither removes the propagation time between a
quantum resource and a remote decision maker. The open problem is therefore not
only how to optimize an allocation, but where to place, distribute and schedule
the act of deciding itself.

\section{Conclusion}

Quantum networks are constrained not only by how entanglement is generated and
stored, but by where and when the network decides how to use it. Because
entanglement decays while control information travels, decision distance enters
delivered fidelity directly. The functional graph makes that cost measurable in
the same multiplicative units as end-to-end state quality. Two geometric limits
follow that should not be collapsed into a generic control overhead: allocation
depends on the distance a request and its decision must travel, which scales with
diameter, whereas reference-free timing depends on propagation between
neighboring repeaters. Controller placement,
decision locality, repeater spacing and slot rate therefore belong beside source
brightness and memory coherence in quantum-network design.

\end{document}